\documentclass[12pt]{article}

\usepackage{times}
\usepackage[figuresright]{rotating}
\usepackage{amsmath}
\usepackage{amsfonts}
\usepackage{url}

\usepackage{natbib}
\usepackage{rotating}
\usepackage{multirow}
\title{Modelling excess zeros in count data: A new perspective on modelling approaches}

\author{John Haslett$^{1}$,
	Andrew C Parnell$^{2}$,
	John Hinde$^{3}$,
	Rafael de Andrade Moral$^{4,*}$
	\\
	\normalsize{$^{1}$School of Computer Science and Statistics, Trinity College Dublin, Ireland}\\
	\normalsize{$^{2}$Hamilton Institute, Insight Centre for Data Analytics, Maynooth University, Ireland}\\
	\normalsize{$^{3}$School of Mathematics, Statistics and Applied Mathematics, NUI Galway, Ireland}\\
	\normalsize{$^{4}$Department of Mathematics and Statistics, Maynooth University, Ireland}\\
	\normalsize{$^*$Rafael.DeAndradeMoral@mu.ie}\\
}

\newcommand {\nc} {\newcommand}

\nc {\Stthfn}[1] {\tilde S_{\theta}\big( #1\big)}
\nc {\Stgmfn}[1] {\tilde S_{\gamma}\big( #1\big)}

\nc {\qA} {q_{_A}}
\nc {\qB} {q_{_B}}
\nc {\qC} {q_{_C}}
\nc {\qD} {q_{_D}}

\nc {\Stththfn}[1] {\tilde S_{\theta\theta}\left( #1\right)}
\nc {\Sthfn}[1] {S_{\theta}\left(#1 \right)}
\nc {\Sththfn}[1] {S_{\theta\theta}\left(#1 \right)}
\nc {\Stomfn}[1] {\tilde S_{\omega}\left(#1 \right)}
\nc {\Stomomfn}[1] {\tilde S_{\omega\omega}\left(#1 \right)}
\nc {\Stthomfn}[1] {\tilde S_{\theta\omega}\left(#1 \right)}
\nc {\Stbetafn}[1] {\tilde S_{\beta}\left(#1 \right)}
\nc {\Stalphafn}[1] {\tilde S_{\alpha}\left(#1 \right)}

\nc \yb {\mathbf{y}}
\nc \Yb {\mathbf{Y}}
\nc \xb {\mathbf{x}}
\nc \zb {\mathbf{z}}
\nc \Xb {\mathbf{X}}
\nc \fb {\mathbf{f}}
\nc \Mb {\mathbf{M}}
\nc \by {\yb}

\nc \EE {\mathrm{I\!E\!}}
\nc \RR {\mathrm{I\!R\!}}
\nc\ZZ {\mathbb{Z}}

\nc \yd {\dot{ y} }
\nc \ydd {\ddot{y}}
\nc {\Yt} {\tilde Y}
\nc {\Ybt}{\tilde {\mathbf Y}}
\nc {\Jt} {\tilde J}
\nc {\kp}{\kappa}
\nc {\pit} {\tilde \pi}
\nc {\pib} {\breve{\pi}}
\nc {\mut} {\tilde \mu}
\nc{\taut}{\tilde \tau}
\nc {\Iy} {\mathbb{I}\{y=0\}}
\nc {\IY} {\mathbb{I}\{Y=0\}}
\nc {\IYi} {\mathbb{I}\{Y_i=0\}}
\nc {\IYik} {\mathbb{I}\{Y_{ik}=0\}}
\nc {\IYt}{\mathbb{I}\{\Yt=0\}}
\nc {\IYbt}{\mathbb{I}\{\Ybt=0\}}
\nc {\IYti}{\mathbb{I}\{\Yt_i=0\}}
\nc {\IYtj}{\mathbb{I}\{\Yt_j=0\}}
\nc {\IYj}{\mathbb{I}\{Y_j=0\}}
\nc {\IYtN}{\mathbb{I}\{\Yt=N\}}
\nc {\IYto}{\mathbb{I}\{0<\Yt<N\}}
\nc {\Iyi} {\mathbb{I}\{y_i=0\}}
\nc {\Iyni} {\mathbb{I}\{y_i \neq 0\}}
\nc {\Iyih} {\mathbb{I}\{y_i=h\}}
\nc{\nz}{\cancel{0}}

\nc{\logit}{\mbox{logit}}
\nc{\expit}{\mbox{expit}}

\nc {\StthYt} {\tilde S_{\theta} (\theta, \gamma; \tilde Y)}
\nc {\SthY}{S_{\theta} (\theta; Y)}
\nc {\Stthy} {\tilde S_{\theta}  \theta, \gamma ; y}
\nc {\Stthyi} {\tilde S_{\theta} ( \theta_i, \gamma_i ; y_i)}

\nc {\Stalphay} {\tilde S_{\alpha} (\theta, \kappa; y)}
\nc {\Stom} {\tilde S_{\omega}}
\nc {\Stq} {\tilde S_{q}}
\nc {\StkpYt} {\tilde S_{\kappa}(\omega \theta ; \tilde Y)}
\nc {\StbetaYt} {\tilde S_{\beta} (\omega, \beta \omega; \tilde Y)}
\nc {\Sbetaj} {S_{\beta_j}}
\nc {\Stbeta} {\tilde S_{\beta}}
\nc {\Stalpha} {\tilde S_{\alpha}}
\nc {\Stgm} {\tilde S_{\gamma}}
\nc {\Stbetaj} {\tilde S_{\beta_j}}
\nc {\Stalphahj} {\tilde S_{\alpha_h}}
\nc {\StbetajYt} {\tilde S_{\beta_j} (\beta, \omega; \tilde Y)}
\nc {\Stbetajy} {\tilde S_{\beta_j} (\omega; y)}
\nc {\StomYt} {\tilde S_{\omega} (\theta, \omega; \tilde Y)}
\nc {\StomhYt} {\tilde S_{\omega_h} (\theta, \omega; \tilde Y)}
\nc {\Stomy} {\tilde S_{\omega} (\theta, \omega; y)}
\nc {\Stth}{\tilde S_{\theta}}
\nc {\Sthy}{S_{\theta}(\theta; y)}
\nc {\Sthyi}{S_{\theta_i}(\theta_i; y_i)}
\nc {\Sthz}{S_{\theta}(\theta; 0)}
\nc {\Sthzi}{S_{\theta_i}(\theta_i; 0)}

\nc{\ddpifn}[1]{\frac{\partial #1}{\partial \pi_0}}
\nc{\ddlmdfn}[1]{\frac{\partial #1}{\partial \lambda}}

\nc{\ddthfn}[1]{\frac{\partial #1}{\partial \theta}}
\nc{\ddkpfn}[1]{\frac{\partial #1}{\partial \kappa}}
\nc{\ddomfn}[1]{\frac{\partial #1}{\partial \omega}}
\nc{\ddgmfn}[1]{\frac{\partial #1}{\partial \gamma}}
\nc {\ddbetajfn}[1] {\frac{\partial #1 }{\partial \beta_j}}
\nc{\ddafn}[1]{\frac{\partial #1}{\partial {\alpha}}}
\nc{\ddahfn}[1]{\frac{\partial #1}{\partial {\alpha_h}}}
\nc{\ddqfn}[1]{\frac{\partial #1}{\partial {q}}}
\nc {\ddkp} {\frac{\partial  }{\partial \kappa}}
\nc {\ddth} {\frac{\partial  }{\partial \theta}}
\nc {\ddthi} {\frac{\partial  }{\partial \theta_i}}
\nc {\ddpiz} {\frac{\partial  }{\partial \pi_0}}
\nc {\ddkpi} {\frac{\partial  }{\partial \kappa_i}}
\nc {\ddkh} {\frac{\partial  }{\partial \kappa_h}}
\nc {\ddom} {\frac{\partial  }{\partial \omega}}
\nc {\ddomh} {\frac{\partial  }{\partial \omega_h}}
\nc {\ddbetaj} {\frac{\partial  }{\partial \beta_j}}
\nc {\ddbetak} {\frac{\partial  }{\partial \beta_k}}
\nc {\ddbeta} {\frac{\partial  }{\partial \beta}}
\nc {\ddalpha} {\frac{\partial  }{\partial \alpha}}
\nc {\dthbetaj} {\frac{\partial  \theta}{\partial \beta_j}}
\nc {\ddomsq} {\frac{\partial^2  }{\partial \omega^2}}
\nc {\ddthsq} {\frac{\partial^2  }{\partial \theta^2}}
\nc {\ddthom} {\frac{\partial^2  }{\partial \theta \partial \omega}}
\nc {\ddombetaj} {\frac{\partial^2  }{\partial \beta_j \partial \omega}}
\nc {\ddbetajk} {\frac{\partial^2  }{\partial \beta_j \partial \beta_k}}

\nc{\grad} {\nabla}

\nc{\FItfn}[1]{\tilde{\mathcal{I}}_{#1}}
\nc{\FIt}{\tilde{\mathcal{I}}_{ \theta, \gamma}}
\nc{\FIti}{\tilde{\mathcal{I}}_{ \omega, \theta_i}}
\nc{\FI}{\mathcal{I}_{\theta}}
\nc{\FIbeta}{\mathcal{I}_{\beta}}
\nc{\FItbeta}{\tilde{\mathcal{I}}_{\beta, \omega}}
\nc{\FItbetom}{\tilde{\mathcal{I}}_{\beta, \omega}}
\nc{\FIomega}{\mathcal{I}_{\omega}}

\date{}

\begin{document} 

\baselineskip24pt
\maketitle 

\begin{abstract}
We consider the analysis of count data in which the observed frequency of zero counts is unusually large, typically with respect to the Poisson distribution. We focus on two alternative modelling approaches: Over-Dispersion (OD) models, and Zero-Inflation (ZI) models, both of which can be seen as generalisations of the Poisson distribution; we refer to these as Implicit and Explicit ZI models, respectively. Although sometimes seen as competing approaches, they can be complementary; OD is a consequence of ZI modelling, and ZI is a by-product of OD modelling. The central objective in such analyses is often concerned with inference on the effect of covariates on the mean, in light of the apparent excess of zeros in the counts. Typically the modelling of the excess zeros \emph{per se} is a secondary objective and there are choices to be made between, and within, the OD and ZI approaches. The contribution of this paper is primarily conceptual. We contrast, descriptively, the impact on zeros of the two approaches. We further offer a novel descriptive characterisation of alternative ZI models, including the classic hurdle and mixture models, by providing a unifying theoretical framework for their comparison. This in turn leads to a novel and technically simpler ZI model. We develop the underlying theory for univariate counts and touch on its implication for multivariate count data.
\end{abstract}

\textbf{Keywords:} hurdle; over-dispersion; zero altered; zero deflation; zero inflation

\section{Introduction}

Regression analysis of count data arises in many fields, including agriculture \citep{Blasco-Moreno2019}, ecology \citep{Mcmahon2017}, climatology \citep{Salter-Townshend2012}, finance \citep{Benson2017}, and pharma \citep{Min2005}. The simplest modelling framework for such analysis is generalised linear modelling using the Poisson and Negative Binomial (NB) families. In regression the mean parameter is related, via a link function, to a linear combination of covariates. In the NB context parameters may also be related to the dispersion parameter.

It is common in such data to encounter an apparent excess of zeros, often with respect to the Poisson and even with respect to the NB. A generic challenge is thus regression of count data in the presence of excess zeros. The mechanism by which these zeros arise may not always be a key focus of the analysis, yet the choice of model for this mechanism may have repercussions on other parameters that are of concern. Our focus is on the alternative models. In common with much of the literature we use the term Zero-Inflated (ZI) rather broadly to refer to several alternatives, although some models necessarily include Zero-Deflation.

The seminal papers are those of \citet{Mullahy1986}, who introduced the `hurdle' models, and of \citet{Lambert1992}, who proposed a mixture model for the zeros. Both can be seen as one-parameter extensions of a simpler base distribution, in which the Poisson (or NB) probability of zero is `altered'. Over the following three decades a very large literature has developed (at the date of writing, about 1900 papers include the terms  `zero-inflated' or `hurdle' in the title; more than 36,000 include the term somewhere in the paper). The issue of dealing with an excess of zeros has become deeply embedded in the methodology of count data regression. 

A complicated nomenclature has developed. Very many authors regard the term ZI to be coterminous with the specific model proposed by Lambert; but other terms used are Zero-Modified, \citep{Dietz2000}, Zero Altered \citep{Yee2015,Rigby2005}, two-part \citep{Pohlmeier1995}, and conditional \citep{Welsh1996}. Various re-parameterisations exist, such as marginal ZI models in which the ZI and Poisson components are merged into a single mean parameter \citep{Martin2017MarginalData, Long2014AEffects}.  Generically we refer to these as explicit ZI models.  Surprisingly, however, there have been relatively few review papers on the topic in the statistical literature \citep[though see][]{Ridout1998,Warton2005,Deng2005,hilbe20077, perumean2013zero, Farewell2017Two-partData}.

A second approach is via distributions that are Over-Dispersed (OD) with respect to the Poisson, the best known example of which is the Negative Binomial. This also is a one-parameter extension of the Poisson. This approach we describe as implicit ZI, for the ZI is a by-product of a wider focus on the mean/variance relationship; in the Poisson distribution the mean and variance are equal. They also can be said to `alter' the probability of a zero from its Poisson `base'. For brevity, below we sometimes refer to the implicit and explicit approaches as OD and ZI respectively.

These two approaches may be combined; zero-inflating the Negative Binomial is then a two-parameter extension of the Poisson. We do not pursue in any detail the many other over-dispersed distributions that have been proposed in recent years, nor their zero-inflated variants, but we remark that this is a rapidly growing  literature. 

But others \citep[eg][]{Warton2005} see them as competitors. In particular, \cite{Warton2005} provided comparisons between different implicit and explicit ZI models fitted to a total of 1672 abundance variables across 20 multivariate datasets. He argued there is little or no evidence for the need to explicitly model excess zeros; the wide class of over-dispersed models is sufficient. However, while his views have not been explicitly rebutted, there are some examples in the literature where explicit ZI is preferred to over-dispersion \citep[see eg][]{Welsh1996,Hall2000} and others where ZI and over-dispersion are used together, typically in a zero-inflated negative binomial (ZINB) model \citep{jansakul2008score}.

This paper examines both approaches with a view to gaining new insights on their different properties. We show that the link between ZI and OD models is complex; both approaches induce specific versions of each other. These insights are primarily theoretical. In particular they suggest reasons why it will often in practice be very difficult to distinguish, from data, which of the alternative models will be `best' in any useful sense. These insights lead to a new form of ZI which we term `Logistic ZI'. 

Some argue that the essential difference between these approaches is that there are situations where at least some of the zeros (and \emph{only} the zeros) arise from a process that differs essentially from that which generates the counts, including counts of zero. This \emph{prescribes} an explicit ZI approach, in which some of the observed zeros are deemed to reflect an unreported structural variable; they may even be deemed `false'. But, although not strictly necessary, this concept of true and false zeros may impose a burden on some users, for an over-dispersed distribution can often achieve the same descriptive effect; see \citet{Blasco-Moreno2019} and \citet{Martin2005}. By contrast, OD distributions can lead to excess low counts \emph{such as} zero. Zeros are no longer special: \emph{low} counts may be inflated with respect to the Poisson. The key distinction is the upper tail of OD distributions is also inflated with respect to the Poisson.

Our fundamental concern in this review is with concepts, rather than with, say, algorithms, power and tests. We defer consideration of vector counts, which introduce new challenges, but not new concepts. We motivate the paper in Section \ref{Sec:motiv_ex}, by referring to a concrete example. We then introduce our general approach including our new type of ZI, in Section \ref{Sec:models_excess_zeros}. Section \ref{sec:ZI_via_NB} examines, from the same perspective, the implicit ZI induced by the Negative Binomial. Our concluding thoughts are in Section 5.

\section{Motivating example}\label{Sec:motiv_ex}

In this section we use a simple dataset to illustrate  a brief and conventional overview of some models that deal explicitly or implicitly with excess zeros in the context of regression. Specifically, we adopt the classic nomenclature and motivation of the models, but defer a discussion of theoretical aspects until Sections 3 and 4. 

The datatset is highly structured, permitting the fitting of 
saturated models for the mean structure. It also has, perhaps somewhat atypically, extensive replication allowing the calculation of natural and non-parametric estimates of, not just, the means but also of the probability of zeros. 
This allows us to see clearly that a basic Poisson regression model deals inadequately with the incidence of zero counts. It also allows us to evaluate the different explicit approaches to the handling of the excess zeros in terms of both the fitting of the probabilities of zeros and the fitting of the means. Both of these are challenging in more general regression settings. But we argue in this paper that the opaqueness, most especially of explicit ZI modelling, is a contributory factor. In the same way we can study implicit ZI models, and here we focus on two versions of the negative binomial regression model as examples of over-dispersed count models.

\begin{figure}[htb]
    \centering
    \includegraphics[width = \textwidth]{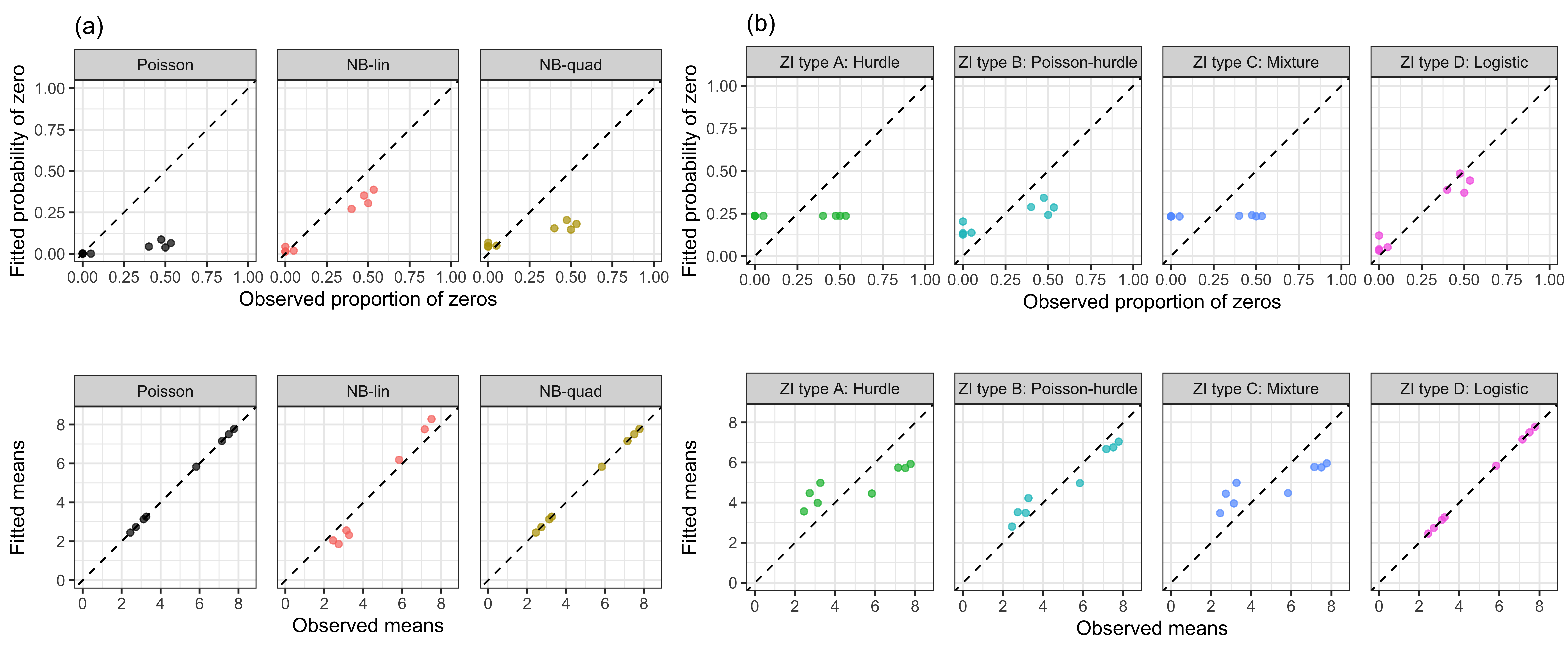}
    \caption{Fitted versus observed proportion of zeros and means of each combination between photoperiod and hormone concentration in the trajan data for (a) the Poisson, NB-lin and NB-quad models, and (b) ZI models types A, B, C and D.}
    \label{fig:trajan_14panels}
\end{figure}

We consider the Trajan apple data of \citet{Ridout1998}. The response variable is the number of roots produced by 270 micropropagated shoots of the apple cultivar Trajan under two different photoperiods (8 and 16 hours) and four different concentrations of the growth hormone cytokinin BAP in a completely randomised design with multiple replication at each of the settings. Here we treat both explanatory variables as factors and restrict attention to the full interaction model for the eight different photoperiod by hormone settings, so in essence fitting the frequency distribution of counts for the replicates at each setting. We refer to these eight distinct settings as $C^{(k)}$, $k=1,\ldots,8$ and index any associated model parameters and data summaries in the same way; writing $\bar y^{(k)}$ and $p_0^{(k)}$, $k=1,\ldots,8$, for the cell means and zero-proportions. Our attention here focusses on the fit of the various models in terms of both the overall means and the zero proportions for each of the eight combinations by comparing the model fitted values with the sample values $\{\bar y^{(k)}, p_0^{(k)}\}$; these are displayed in Figure~\ref{fig:trajan_14panels}. Each of the seven different models shown in Figure~\ref{fig:trajan_14panels} are explained below, and we encourage the reader to read this section whilst repeatedly returning to Figure~\ref{fig:trajan_14panels} for reference.

We start with the basic Poisson model with $\pi_y(\lambda)=\lambda^y e^{-\lambda}/y!,\, y\in \mathbb{N}$. Fitting a Poisson log-linear full interaction model corresponds to estimating a separate parameter $\lambda^{(k)}$ for each of the settings $C^{(k)}$ with the cell means $\bar y^{(k)}$ being the maximum likelihood estimators.  Effectively the data are treated as iid, within each cell. So here the observed cell means are reproduced by the fitted model (see Figure~\ref{fig:trajan_14panels}(a), bottom-left panel), which can be viewed as a consequence of the simple Exponential Family form of the Poisson. For the zero proportions the maximum likelihood  estimators are $\hat\pi_0^{(k)}=\exp(-\hat\lambda^{(k)})=\exp(-\bar y^{(k)})$ are not unbiased, unlike the cell zero-proportions $p_0^{(k)}$ which are non-parametric and unbiased estimators of $\pi_0^{(k)}$. The comparison of these in Figure~\ref{fig:trajan_14panels}(a) shows the a priori evidence of excess zeros compared to a Poisson model, especially for the 16 hour photoperiod where there are many observed zeros along with some large non-zero counts. This motivates the need to explore ZI extensions of the basic Poisson model.

\subsection{Explicit ZI extensions}

Here we revisit the two classical approaches: the hurdle model of Mullahy, in two variants, and the mixture model of Lambert. In Section 3 these will be referred to as types A, B and C, and a fourth Type D will be introduced. Results from all four are presented in Figure~\ref{fig:trajan_14panels}(b) where comparisons are made with the aid of the non-parametric estimates $\{\bar y^{(k)}, p_0^{(k)}\}$. 

We denote these explicit ZI extensions  by $\tilde\pi_y(\lambda,\gamma)$, where here we focus on one parameter extensions  of the underlying Poisson base model $\pi_y(\lambda)$. Section 3 will provide a unified treatment of the way that $\gamma$ can be seen as parameterising different types of explicit ZI. In the brief overview below of the classical treatment of hurdle and mixture models, $\gamma$ implicitly parameterises terms specific to these models. The key step is that $\pit_y$ arises from a specific alteration of the Poisson $\pi_0$ to $\pit_0$; and that, for $y>0$, $\pit_y = \rho \pi_y$ where $\rho$ re-normalises to ensure that $\sum_y \pit_y = 1$. So for $y>0$ the modified distribution retains the same structure as the Poisson for successive terms, i.e. $\pit_{y+1}/\pit_y= \pi_{y+1}/\pi_y$.

The first class of ZI extensions that we consider are the Hurdle models of \citet{Mullahy1986}. The basic idea of the hurdle model is that the zeros and non-zeros arise from separate processes, allowing the zeros to be modelled as binary outcomes and the non-zero counts as outcomes from a zero-truncated count distribution. The hurdle name comes from the idea that the binary process for the zero/non-zero values essentially models the probability of crossing over the hurdle from zero to non-zero counts. Then conditional on having crossed over this hurdle the model is a standard count data model for the non-zero counts, typically taken to be a zero-truncated  count model. The simplest hurdle model is one in which the zeros are modelled as coming from a Bernoulli distribution with a constant probability $\pit_0 = \pit_0(\gamma)$ and the non-zero counts coming from a zero-truncated Poisson distribution; we refer to this in Section 3 as Type A. The zero-truncated Poisson distribution has mean $E[Y | Y>0] = \frac{\lambda}{(1-e^{-\lambda})}$. 
In our notation this hurdle model has the form
\begin{equation} \label{ZI eq}
\pit_y(\lambda, \gamma) = 
\begin{cases}
			\pit_0&
			\text{ if } y=0\\
		\displaystyle	\frac{(1-\pit_0)}{(1-e^{-\lambda})}\, \pi_y(\lambda) &\text{ if } y>0
		\end{cases}
\end{equation}
with the re-normalising $\rho=\frac{(1-\pit_0)}{(1-e^{-\lambda})}$ and $E[Y] =\frac{(1-\pit_0)}{(1-e^{-\lambda})}\lambda =\rho\lambda$.

Fitting this simple model to the Trajan data with a single additional ZI parameter, the common $\pit_0$, with separate parameters $\lambda^{(k)}$ for each cell $C^{(k)}$, the overall proportion of observed zeros ($0.237$) is necessarily reproduced by the fitted model. For under this model, the indicator variable $E[\IY] = \pit_0$ irrespective of cell and is estimated by the overall proportion of zeros $p_0$. But as the eight individual cells have different numbers of zero observations, $\pit_0 \ne p_0^{(k)}$ in general; we elaborate in Section 3. As the zero-truncated Poisson distribution is in the Exponential Family, the fitted model reproduces the means of the zero-truncated data; thus $E[Y^{(k)}|Y^{(k)}>0]$ is estimated by the truncated mean $\bar y_+^{(k)}$ for each cell, and hence the overall fitted values, $(1-\hat\pit_0)\bar y_+^{(k)}=(1-p_0)\bar y_+^{(k)}$, differ from the sample values $\bar y^{(k)}$ as seen in Figure~\ref{fig:trajan_14panels}(b).  

More general hurdle models, in particular where $\pit_0$ is itself modelled via covariates in a regression context, involve specifying a particular form for the binary model. In addition to any specific linear predictor, this specification includes the choice of a link function for the zero probability $\pit_0$, or equivalently for the probability of non-zero counts $\pit_+=1-\pit_0 $, where $\pit_+$ can be thought of as the probability of crossing the hurdle. 
The link function can be taken as any standard binary/binomial model link but can also be motivated by considering the zeros to have come from a particular (truncated) count distribution; for example, \cite{Mullahy1986} considers the use of the Poisson distribution leading to a complementary log-log link for the hurdle crossing probability $\pit_+$ .

In the general form of Mullahy's Poisson-hurdle the zero and non-zero counts are both assumed to come from Poisson distributions but with different (modelled) parameters. A simplified version of this, with a single additional ZI parameter, takes the Poisson parameters as proportional with $\pit_0(\lambda, \alpha)=\pi_0(\alpha \lambda)$ with $\alpha<1$ for zero-inflation and where $\alpha=1$ reduces to a common Poisson model for zero and non-zero-counts (to fit in with our general notation for $\pit_0(\lambda, \gamma)$ in Section 3 we take $\alpha=e^{\gamma}$) ; we refer to this as Type B. In this shared parameter version of the hurdle model we no longer have the separation of the zero and non-zero processes and consequent simplification of the estimation. Here both zero and non-zero counts contribute to the estimation of the regression parameters and not surprisingly this model does better than Type A at reproducing the overall means and also manages to pick up some of the differences in zero proportions.

The other broad class of explicit ZI approaches are the zero-inflated mixture models, exemplified by the zero-inflated Poisson (ZIP) model of \citet{Lambert1992}. Here the zero probability is extended to a mixture of the base zero probability and degenerate point mass at zero, where the mixing probability provides the additional parameter and corresponds to the specific inflation of the zero probability. In the usual mixture-like notation we have $\pit_0(\lambda, q) = q + (1-q)\pi_0$ for $0\le q\le 1$, with $\rho=1-q$. This can be written in our generic extended form $\pit_0(\lambda, \gamma)$ by parameterizing $q$ as a function of $\gamma$; specifically, for reasons that will become clear in Section~3, we take $q=1-e^\gamma$ giving $\gamma=\log(1-q)$. We refer to this as Type C. For the Trajan data we use a constant mixing probability $q$ (equivalently constant $\gamma$) and separate $\lambda^{(k)}$ values for each of the eight classes $C^{(k)}$. From Figure~\ref{fig:trajan_14panels}(b) we see that here the results are very similar to those for Type A, this is because the cell means are generally large and so the base Poisson model zero probabilities, $\exp(-\hat\lambda^{(k)})$, are small and contribute little to the fitted zero probabilities that are dominated by the estimated constant zero inflation $\hat{q}$.

As a further novel explicit ZI model in Section~\ref{Sec:models_excess_zeros} we introduce Type D, where the alteration of the zero probability is made through its  odds-ratio. As discussed in Section 3, the resulting extended distribution is in the Exponential Family and for the Trajan data interaction model with the eight distinct $\lambda^{(k)}$ and a single ZI parameter the sufficient statistics are the individual cell means, $\bar y^{(k)}$, and the overall proportion of zero counts. Consequently, the maximum likelihood fitted model  reproduces the eight cell means and the overall proportion of zeros and, here, also seems to recover much of the individual cell  zero probability structure, all with a single additional parameter.

While we would not claim that the pattern of behaviour across Types A -- D seen with the Trajan data is going to hold in general, we do feel that there are some important messages. First, the choice of ZI model may have unforeseen consequences on the overall fit, although the specific ways in which this may happen are rather subtle. Second, mis-specification of the ZI parameter (as here in treating it as constant when then is a large difference across the two photoperiod settings) can severely impact the estimation of the mean component of the model and covariate parameters of interest as seen here for types A and C.

Note that for the Trajan data  if a full interaction model is also used for the additional ZI parameter, then all four types (A, B, C, and D) are equivalent, as the likelihood  reduces to one based on eight independent samples with their own parameters. These comments suggest that in practice if interest is in the mean model then it would be sensible to use a rich model for the ZI parameter, which will provide more robust inferences. Of course, if there is also explicit interest in the covariates affecting the additional zero part then some care may be needed in choosing an appropriate model and the development of some targeted diagnostics to help in this would be valuable.

\subsection{Implicit ZI extensions}

We now consider the alternative approach of using over-dispersed models with apparently zero-inflated data. In general, many over-dispersed models are one-parameter extensions of the basic Poisson model and we write their probability function as $\pi_y(\lambda,\phi)$ with an additional dispersion parameter $\phi$, where often $\phi=0$ corresponds to the Poisson. Here we restrict our attention to fitting two different forms of negative binomial model. The canonical negative binomial model arises from a Poisson-gamma mixture taking $Y\sim\text{Poisson}(\lambda V)$ where $V$ is a Gamma$(1/\phi,1/\phi)$ distributed random variable with mean $1$ and variance $\phi$. The resulting negative binomial distribution, which we refer to as NB-quad, has mean $\mu=\lambda$ and a quadratic variance function $\mu+\phi\mu^2$ and for a fixed value of $\phi$ it is in the Exponential Family. The probability of a zero observation is $\pi_0(\lambda,\phi)=(1+\phi\lambda)^{-\phi^{-1}}\ge e^{-\lambda}$ with equality for $\phi=0$, showing that, for $\phi>0$, this model does indeed inflate the probability of a zero observation compared to the Poisson distribution. For the Trajan data fitting the NB-quad with the full eight parameter interaction model and a single dispersion parameter $\phi$, we see from Figure~\ref{fig:trajan_14panels}(a) that this model also reproduces the cell means and hence has some robustness for inference on regression parameters. However, while the single additional over-dispersion parameter accounts for some of the zero-inflation, it fails to model the zero proportions well. Of course, as with the explicit ZI models, by including a model for the over-dispersion parameter we can substantially improve the fit of the zeros, at the price of added complexity. 

For comparison we also consider a different parameterization of the negative binomial model (arising from a different version of the Poisson-gamma mixture) that has a linear variance function, $\mu (1+\phi)$ and which we refer to as NB-lin. As with all Poisson mixture models, as well as over-dispersion, the probability of zero is inflated compared to the Poisson with now $\pi_0(\lambda,\phi)=\left((1+\phi)^{\phi^{-1}}\right)^{-\lambda}\ge e^{-\lambda}$, again with equality for $\phi=0$, This model is not in the Exponential Family and even when fitting the full eight parameter interaction model for the $\lambda^{(k)}$ it does not reproduce the cell means, see Figure~\ref{fig:trajan_14panels}. However, it does seem to recover more of the structure in the zero proportions. The point here is that different over-dispersed count models may perform better, or worse, in particular examples. Of course, there are many other different over-dispersion models that we could consider, but none are going to be a panacea for fitting all zero-inflated data.

The intention here is not to present a definitive analysis, but rather to consider a simple common model where differences are apparent between the different implicit and explicit ZI models. In particular, we restrict our attention to a single ZI or over-dispersion parameter. In practice, as suggested above, we may want this parameter to depend on covariates and then many of the ZI models can be very similar, or indeed identical, and over-dispersion models can also give similar fits, as noted by \citet{Warton2005}. 

Here for explicit ZI models we have taken the base model to be Poisson, but this is not necessary and we could use any suitable count distribution that could itself be a one (or more) parameter extension of the Poisson. Such combinations of explicit and implicit approaches lead to two parameter extensions $\pit_y(\lambda,\phi,\gamma)$ of the Poisson model and offer yet more flexibility. Additionally, in principle, it is also possible to have regression models for each of the three parameters. However, such flexibility and complexity comes at a price and it would require a very rich dataset to distinguish between different models.

\section{Explicit Models for Excess Zeros}
\label{Sec:models_excess_zeros}

Here we consider formally the explicit ZI modelling of excess zeros in count data regression. We revisit and generalise the analysis of the Trajan data. Our objective in this is to put extant models into a theoretical framework which seems to be novel. This emphasises the functional form of what was described above as the specific alteration of the Poisson $\pi_0$ to $\pit_0$. This apparently new perspective may facilitate the (under-developed) constructive criticism of model fit, generally, using the Trajan analysis as a primitive example. Further, a new ZI type emerges naturally from the theory, which can inherit the attractive properties of Exponential Family probability distributions.

Formally, we discuss a wide family of one parameter extensions of the count distribution $\pi_y(\lambda)$ underlying a generalized linear model (GLM), leading via coefficients to new estimates of the mean and thus of the probability of zero counts. The extensions are based on a new model $\pit_y$ differing from the \emph{base} $\pi_y$ via a  parameter, here termed $\gamma$. This itself could  be modelled as depending on covariates via further coefficients. This is indeed routine, but is not our immediate concern. Here, by studying a very wide range of functional options for $\pit_y(\lambda,\gamma)$, we hope to assist in the search for parsimonious ZI alternatives to the base pmf. 

We first formalise the modelling in the previous section. The immediate focus is with models which \emph{explicitly} address the apparent excess of zeros, but only zeros. The base could in fact be provided by one of many two (or more) parameter over-dispersed pmfs, denoted here by $\pi_y(\lambda, \phi)$; often these are themselves generalisations of the Poisson. But for simplicity of notation, we typically suppress the second parameter. Here we refer to a zero-altered distribution $\pit_y$ as defined by a function $\pit_0(\pi_0, \gamma)$ with consequent implications for $\pit_y, y \ne 0$. As we develop below the functional form defines a type of ZI, parameterised in the notation below by $\gamma$. Three such types (A,B,C) were discussed in Section 2 and a fourth (type D) mentioned. 

The \emph{a priori} need for such an extension was established there by a comparison of proportions $p_0^{(k)}$ with fitted $\hat{\pi}_0^{(k)} = \pi_0(\hat \lambda^{(k)})$, where $\pi_y$ denoted a Poisson GLM, leading to the maximum likelihood estimate for each $\lambda^{(k)}$. These plots re-appear in Figure~\ref{fig:trajan_pit0_pi0}. In each of the panels, the Poisson GLM is represented by the diagonal; the points are pairs $\left(\hat{\pi}_0^{(k)},p_0^{(k)}\right)$. Note now that the horizontal axis in Figure~\ref{fig:trajan_14panels} denotes $\pit_0$ for the given model whereas in Figure~\ref{fig:trajan_pit0_pi0} it denotes the (Poisson) base $\pi_0$. Note also the full unit square is used for the plots, in contrast to Figure~\ref{fig:trajan_14panels}, to allow for  examination of the functions underpinning explicit ZI modelling. As in the previous section, a full explanation of these functional forms is provided below and should be read in conjunction with Figure~\ref{fig:trajan_pit0_pi0}. Furthermore the OD models, which are also presented in Figure~\ref{fig:trajan_pit0_pi0}, should be read in conjunction with Section \ref{sec:ZI_via_NB}.

The details of these generalisations are the subject of this Section, as we now develop, firstly focussing on the theory of the ZI types $\pit_y$, and subsequently by considering the implications for likelihood inference.

\subsection{Explicit ZI Types}

Formally, recalling that $\pi_y(\lambda)$ refers to an arbitrary base pmf, with $E_{\pi}[Y]=\lambda$ and and possibly other parameters $\phi$, we write
 \begin{equation} \label{ZI.3 eq}
 \pit_y(\lambda, \gamma) = 	\begin{cases}
 			\pit_0(\pi_0, \gamma)&
 			\text{ if } y=0\\
 			\rho(\pi_0,\gamma) \pi_y(\lambda) &\text{ if } y>0
 		\end{cases}
 \end{equation}
 where $\rho$ re-normalises. The requirement that $\sum_y \pit_y = \sum \pi_y=1$ leads to  $\rho = \frac{1-\pit_0}{1-\pi_0}$. Note that, under $\pit_y, E_{\pit}[Y] = \mu = \rho \lambda$ and so the mean is also altered. 
 
The specific function $\pit_0(\pi_0, \gamma)$ of $\pi_0$ characterises the \emph{type} of ZI  and is parameterised by $\gamma$; it is such functions that are illustrated in Figure~\ref{fig:trajan_pit0_pi0}.  Typically, but not necessarily, $\gamma = 0$ denotes the null case $\pit_0=\pi_0$. Observe that parameters $(\lambda, \phi)$ enter $\pit_0$ and $\rho$ only \emph{via} $\pi_0$.  Note also that the \emph{relative} magnitudes of  $\pit_y$ for non-zero $y$ are the same as those for the base $\pi_y$.  Equivalently, the conditional pmfs, given $y > 0$ (that is, the zero-truncated distributions), are the same, whether based on $\pit_y$ or the base $\pi_y$. In particular $\mu^+ = E_{\pit}[Y|Y>0] = E_{\pi}[Y|Y>0] = \lambda^+$. It is such conditional pmfs that distinguish explicit and implicit approaches to the modelling of excess zeros. 

We now consider special cases. Three are relatively common. We will see below that many of the interesting  cases can be written as
 \begin{equation} \label{gform}
g(\pit_0)=\gamma +g(\pi_0).
\end{equation}
They are all summarised in Table~\ref{tab:functions}.

\subsubsection*{ZI Type A} The simplest model has constant $\pit_0 \in [0,1]$. With this restriction we could write $\pit_0 = \gamma$. But here, as for other types, $\gamma$ may of course depend on covariates; we thus write $g(\pit_0) = \gamma$ for some link function $g(.)$. In this  form $\gamma$ is typically unconstrained.

This form is the hurdle model discussed in Section 2; it could be described as the \emph{constant} type of ZI. But its use in this very basic form (constant $\pit_0$) is almost degenerate.  Its inadequacy for the Trajan data is apparent (Figure~\ref{fig:trajan_pit0_pi0} top right panel). Note that here, and equivalently in the other panels, the value of $\gamma$ used for the function is the MLE $\hat \gamma$ for the Trajan data under Type A, as discussed in Section 2. In this case it is that value of $\gamma$ that renders $\pit_0 = p_0$, the overall proportion.  Further, for this and other panels, the `data points' are pairs $( \pi_0(\hat \lambda^{(k)}), p_0^{(k)})$ under each ZI type. They thus correspond to the plotted functions $\pit_0(\pi_0(\lambda), \hat \gamma)$, where the value of $\hat \gamma$ is specific to the ZI type.  For Type A, the MLE $\hat \lambda^{(k)}$ is derived from the truncated mean $\bar y_+^{(k)}$, as previously discussed. We formalise such inference in the next subsection.

Observe that this ZI type is zero-inflationary for small $\pi_0$, but zero-deflationary for large $\pi_0$. This is a necessary consequence of its motivation, where zero counts are generated by a process quite independent from positive counts. Note also that the neutral case of $\pit_y = \pi_y$ is \emph{not} a special case of this model.

\subsubsection*{ZI Type B} 
 
Type B is that referred to in Section 2 as the Poisson-hurdle. It can be expressed as $\pit_0=\pi_0(\alpha\lambda)=(\pi_0)^{\alpha}$ for positive $\alpha$; for the Poisson base model we have $\pit_0=e^{-\alpha\lambda}$ i.e. coming from a Poisson model with mean $\alpha\lambda$. An alternative form is through the complementary log-log link function with unconstrained additional parameter $\gamma=\log(\alpha)$ and $$\log(-\log(\pit_0)) = \gamma + \log(-\log(\pi_0)),$$ giving $\pit_0= (\pi_0)^{e^\gamma}$. Observe that $\gamma =0$ defines the neutral case, with positive and negative $\gamma$ corresponding to under- and over-inflation. The ZI Type B panel in Figure~\ref{fig:trajan_pit0_pi0} (bottom left panel) makes it quite clear that this more clearly reflects the frequency of zeros observed in the Trajan data.

\subsubsection*{ZI Type C}

In the notation of Section 2 this can be written as $(1- \pit_0)=(1-q)(1-\pi_0)$ for $0 \le q \le 1$; zero inflation is thus non-zero deflation by a constant factor. It could be characterised as the `linear model'. It is very widely used; indeed for very many authors, this model is coterminous with ZI. For the Trajan data, like Type A, it is clearly a poor model of the observed zeros (Figure~\ref{fig:trajan_pit0_pi0} middle bottom panel).

Its simplest link-function expression is via the complementary log function for $\gamma \le 0$.
$$
\log(1-\pit_0) = \gamma + \log(1-\pi_0)
$$
It is apparent that for Type C $\log(\rho) = \gamma$. When $\gamma =0$, $\pit_0=\pi_0$ for all $\pi_0$; here, as in Type B, the neutral $\pit_0 = \pi_0$ is a special case.  

Like types B and D, and in contrast to A, $\pit_0 =1$ when $\pi_0 = 1$. But like Type A, when $\pi_0 = 0$, $\pit_0 > 0$; this distinguishes itself qualitatively from Type B, and also from D below.  This is the feature which marks both types A and C as inappropriate single parameter extensions for the Trajan data. Note also that the model is also defined for $\gamma >0$, subject to $\pit_0 > 0$. Thus Type C includes under-inflation; but now $\gamma$ is constrained by a function of $\lambda$. More formally, when $\gamma >0$, we must write $\pit_0 = \max(0,1-e^{\gamma}(1-\pi_0))$. 

\subsubsection*{Type D}

Type D, which is new, is most simply expressed as $$\logit(\pit_0)=\gamma + \logit(\pi_0)$$ for unconstrained $\gamma$. Equivalently we have $\frac{\pit_0}{1-\pit_0}=e^{\gamma}\frac{\pi_0}{1-\pi_0}$. The alteration may now be seen as multiplicatively altering the odds ratio of a zero count.  In closed form, it may be written as $\pit_0 =\frac{e^{\gamma}\pi_0}{1+(e^{\gamma}-1) \pi_0}$.

Although its performance is similar to Type B, arguably it is the best of the four ZI types in describing the observed frequency of zeros in the Trajan data (see Figure~\ref{fig:trajan_pit0_pi0} bottom right panel). But it also differs from the others in a more theoretical sense, as we elaborate below. For, in many cases, pmfs involving Type D extensions are often within the Exponential Family. It is for this reason that D is, for the Trajan analysis, the only one of the four ZI types for which $\hat{\mu}^{(k)} = \bar {y}^{(k)}$, as seen in Figure~\ref{fig:trajan_14panels}, sharing this property with the Poisson and NB-quad distributions.

We conclude this subsection by making two remarks. Firstly, the use of the unit square for the plots of $\pit_0(\pi_0,\gamma)$ emphasises a peculiar feature of the design of the Trajan data. It is  lacking any samples corresponding to small means $\lambda$  and thus large $\pi_0$. It is this that renders it difficult to distinguish which is the `best' model, despite its other peculiar feature -- replication -- that gives access to conditionally iid data and thus to the proportions $p_0$ of observed zeros. An even more extreme example would be fully iid data. There we would see that all the four ZI types would in fact be re-parameterisations of each other and would thus be impossible to distinguish. All four functions would intersect at $\pit_0 = p_0$.

Figure~\ref{fig1_theoretical} uses this to provide a different contrast of the functions. There it is seen that types B and C can be remarkably similar for mid to large values of $\pi_0$, and thus might be difficult to distinguish in a data design that did not include small values of $\pi_0$. Further, Type D distinguishes itself from Type B in two ways. Firstly it has a symmetry (around the diagonal $\pit_0 = 1-\pi_0$) that the others lack; but this may in fact be disadvantageous in some cases. For example, it can be steeper than B for very small $\pi_0$; but then, necessarily, it will be much flatter for small $1-\pi_0$.

Secondly we observe that there are many more options in the design of even the one parameter functions $\pit_0=\pit_0(\pi_0,\gamma)$ than the literature might suggest. These clearly extend beyond the four types we have used for illustration. They typically involve classic link functions $g(p)$, such as complementary log-log and logit. Interestingly however, the particular characteristic of the overwhelmingly popular Type C -- namely that $\pit_0>0$  when $\pi_0 = 0$ -- seems not to be a property associated with any popular link function other than complementary log-log or close but uncommon relatives such as $g(p)=\frac 1 {1-p}$. Furthermore they may all be used with over-dispersed base pmfs $\pi_y(\lambda, \phi)$.

We now address issues in inference.

\subsection{Inference}

We focus on likelihood inference, and illustrate  with the analysis of the Trajan data in Section 2, which provides some new insight. We do not dwell on algorithms or second-order issues; but it may be that our perspective will stimulate ideas on the critical evaluation of explicit ZI modelling. The theory thus focusses on the score equations, being differentials of $\log L = \sum_i \log(\pit_{y_i}) = \sum_i \ell_{y_i}(\mu(\lambda,\gamma), \gamma)$, wrt the parameters $\gamma$, $\lambda_i$ or functions of these such as $\pit_0$, $\mu_i$, and/or the coefficients $\beta$ underlying the $\lambda$ terms. This extends of course to cases of $\pi_y$ where the $\phi$ parameter is non-null; but we do not dwell on this.  Recall that, for the Trajan model, there are nine parameters: one common $\gamma$ term and eight values of $\lambda^{(k)}$, or equivalently of $\mu^{(k)}$. Particular interest lies in useful decompositions of the likelihood, as we now discuss.

\begin{figure}[htb]
    \centering
    \includegraphics[width = \textwidth]{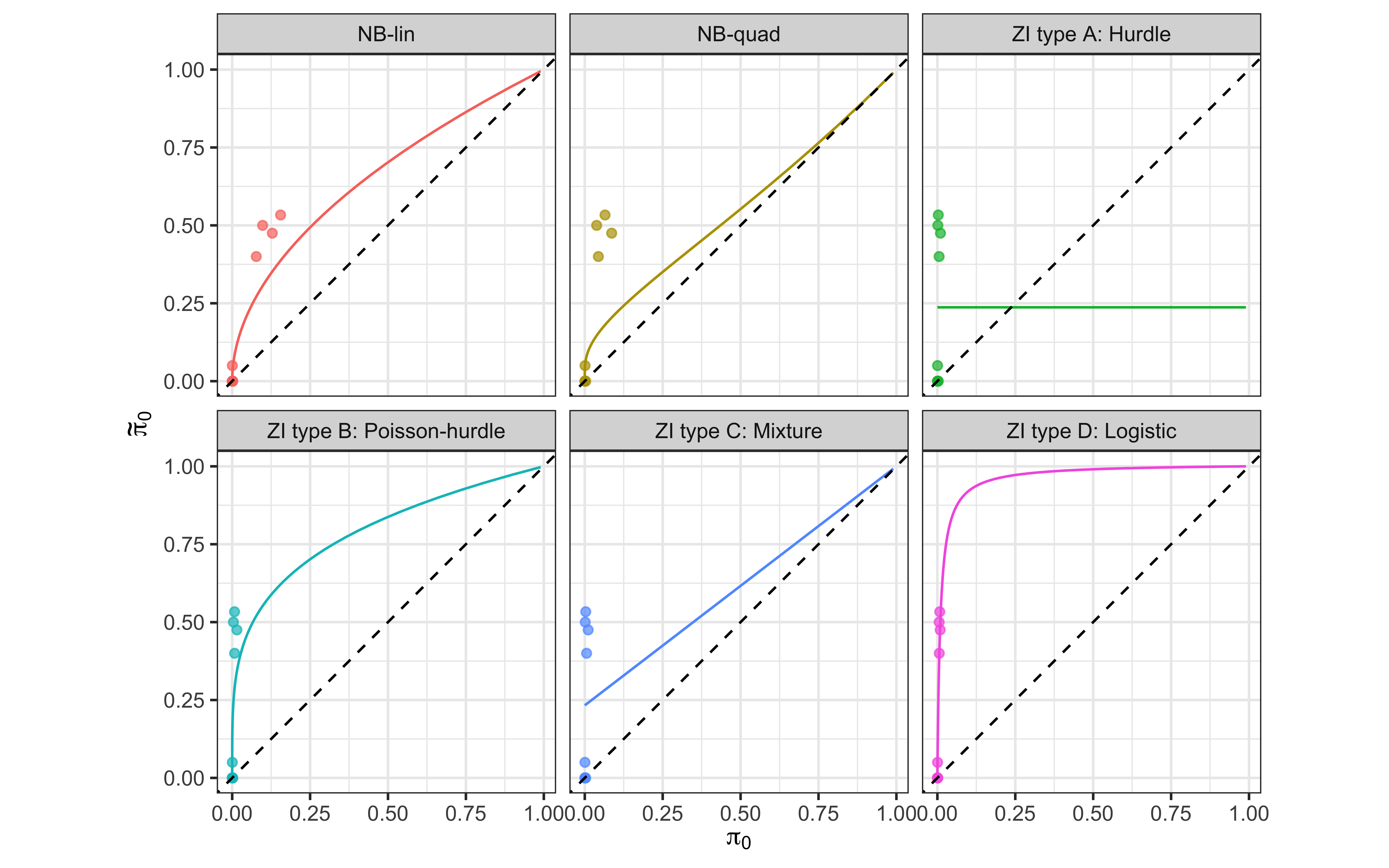}
    \caption{Altered probability of zero $\pit_0$ versus probability of a zero based on the Poisson distribution $\pi_0$ for the NB-lin, NB-quad and ZI models types A, B, C and D, fitted to the Trajan data. Points correspond to the pairs $\left(\hat{\pi}_0^{(k)},p_0^{(k)}\right), k=1,\ldots,8$, i.e., the eight combinations between photoperiod and hormone concentration. The dashed identity line is the Poisson base.}
    \label{fig:trajan_pit0_pi0}
\end{figure}

\begin{figure}[htb]
	\centering
	\includegraphics[width=\textwidth]{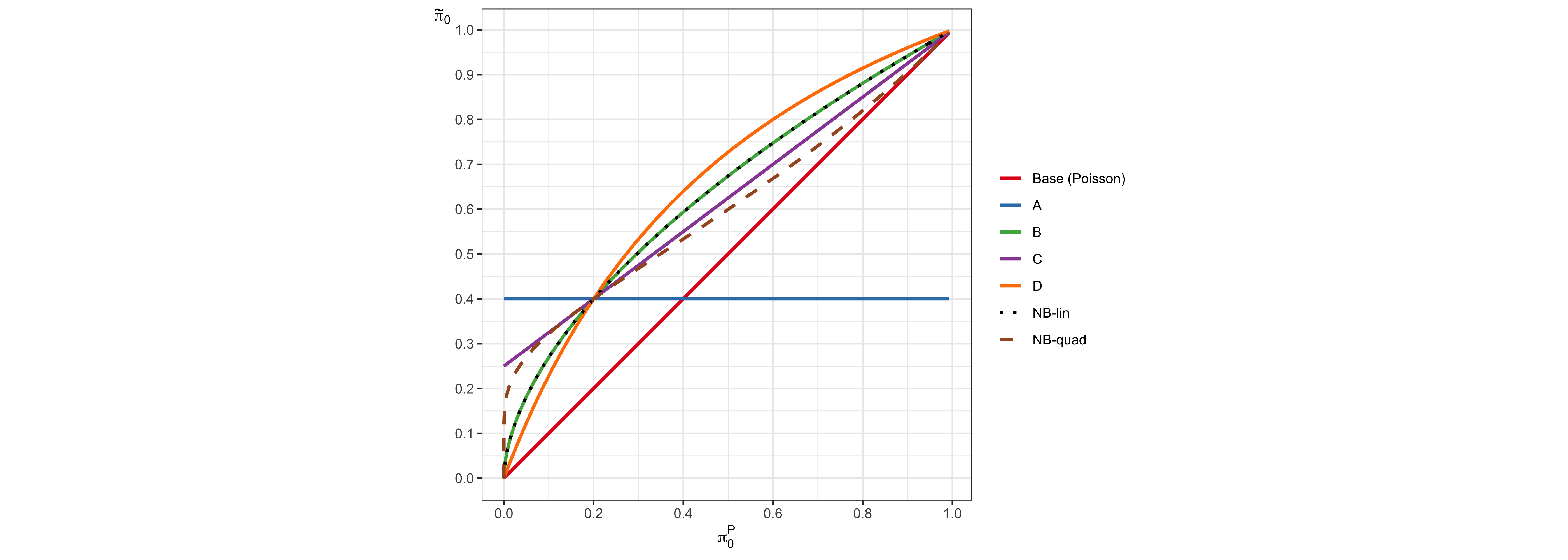}
	\caption{Theoretical ZI relationships. We plot $\pi_0^P$ as Poisson against $\pit_0$ for the ZI and OD types defined in column 1 of Table \ref{tab:functions}. The parameters of each distribution are matched so that each line (except for Poisson) travels through the point (0.2, 0.4) for a range of means. This yields differing ZI parameters for each distribution: $\gamma_A = -0.405$, $\gamma_B = -0.563$, $\gamma_C = -0.288$, and $\gamma_D = 0.981$, all provided to match the scale of Table~\ref{tab:functions}. For the Negative Binomial models $\phi_{\mbox{NB-lin}} = 1.82$ and $\phi_{\mbox{NB-quad}} = 1.13$, again with reference to the parameterisation in Table~\ref{tab:functions}.}
	\label{fig1_theoretical}
\end{figure}

The full distribution $\pit_y(\lambda, \gamma)$ can be written in various forms some of which lead to different decompositions of $\log L$. Although all forms apply to all ZI types, these are most simply interpreted for types A and D.   Equation \eqref{inf1} below decomposes into components of the likelihood focussing on zero and non-zero counts. It sheds light on the fitted functions in Figure~\ref{fig:trajan_pit0_pi0} for all types, and on the $(\lambda,\gamma)$ parameters in Type A. Equation \eqref{inf2} sheds light on Type D, for this is seen often to lie within the Exponential Family, with consequent other decompositions.

We begin by noting alternative versions of \eqref{ZI.3 eq} which are particularly insightful
\begin{align}
\pit_y & = \left(\pit_0\right)^{\Iy}\left(\rho\pi_y\right)^{1-\Iy}=\left(\pit_0\right)^{\Iy}\left(\frac{1-\pit_0}{1-\pi_0}\pi_y\right)^{1-\Iy}\nonumber\\
 &=\left((1-\pit_0)^{1- \Iy}\pit_0^{\Iy} \right)
  \left( \pi_y^+ \right)^{1- \Iy} \label{inf1}\\
& = \left( \frac{\pit_0}{\rho \pi_0}\right)^{\Iy} \rho\pi_y = e^{\gamma \Iy}\left( \frac{ \pi_y}{ \pi_0}\right)e^{-\gamma}\pit_0,  \label{inf2}
\end{align}
where $\pi_y^+ = \frac{\pi_y}{1-\pi_0}$ is the zero-truncated form of $\pi_y$ and $e^{\gamma}$ denotes the odds ratio for $(\pit_0,\pi_0)$. From \eqref{inf1}, with an obvious notation
$\log \pit_y(\lambda, \gamma) = \ell_y^0(\pit_0(\pi_0,\gamma))+ \ell_y^+(\lambda)$, the log-likelihood decomposes as:
\begin{equation}\label{logL}
\log L = \sum_i \ell_{y_i}^0(\pi_0, \gamma)+ \sum_i \ell_{y_i}^+(\lambda_i).
\end{equation}
This result applies to any ZI type, including A -- D above. Importantly the second term is independent of $\gamma$. The first term depends on  $\Iyi$ and  supplies the only information on $\gamma$ through $\pit_0$; but recall that, apart from Type A, $\pit_0$ is also a function of $\lambda$ through $\pi_0$, and thus the first term generally supplies some information on $\lambda$.  The usual calculus leads to score functions whose solution is the MLE for $(\lambda_i, \gamma)$ (and for any other parameters, $\phi$). In general regression, the $\lambda_i$ are dictated by coefficients, these and $\gamma$ being the true target of the inference. 

In the case of Type A, of course,  $\pit_{0i}$ is independent of $\hat {\pi}_{0i}$, this being  $e^{-\hat {\lambda}_i}$ for the Poisson base. In the case of iid observations, as within a single cell of the Trajan design, $\hat \gamma$ is such that $\pit_0(\hat \gamma) = p_0$ the observed proportion of zeros in the data. Recall that this remark applies to \emph{all} two parameter $(\lambda, \gamma)$ types of ZI in the iid case. However, in the Trajan case with common $\gamma$ across all cells, results will differ with  ZI type, as we now illustrate.

It is useful first to consider in more detail the estimation of $\lambda$ under Type A, first for the iid case and then for the Trajan design. For Type A, from \eqref{inf1}, the estimation of $\lambda$ focusses solely on $\ell_y^+(\lambda)$, and thus on the zero-truncated distribution $\pi_y^+(\lambda)$. The expected value of the zero-truncated distribution is $(1-\pi_0)^{-1}\lambda$ for which the MLE is $\bar {y}^+$, this being the case for all pmfs in the Exponential Family, such as the Poisson and NB-quad, and truncated versions thereof. Thus $\hat \lambda = (1-\hat{\pi}_0)\bar y^+$. But $E_{\pit}[Y] = \mu = \rho \lambda$; it is thus estimated, for Type A, by $\hat \mu =\hat \rho \hat \lambda = \frac{1-p_0}{1-\hat \pi_0}(1-\hat \pi_0)\bar {y}^+ = \bar y$. Again, since all two parameter ZI types are equivalent in the iid case, this result applies also to them, as is well known. We revisit this for Type D below, where this demonstration is very simple.

But in the Trajan model, there are eight cells with $\lambda=\lambda^{(k)}$ in each but sharing a \emph{common} ZI parameter $\gamma$, defining for all ZI types, a common $\pit_0$ estimated by the overall $p_0$. The term $\ell_y^+$ decomposes into sums over each of these cells, each leading, separately, to estimates of each $\lambda^{(k)}$. In the case of Type A, these are zero-truncated means and thus lead to estimates  $$\hat\mu^{(k)}=\hat\rho^{(k)}\lambda^{(k)}=(1-p_0)\bar{y}^{+{(k)}} \ne (1-p_0^{(k)})\bar{y}^{+{(k)}} =\bar{y}^{(k)}.$$  It is this that leads, in Figure~\ref{fig:trajan_14panels}, to the fact that  ZI Type A leads to values $\hat\mu^{(k)}$ other than $\bar y^{(k)}$, unlike the simple Poisson. Similar but technically more difficult arguments apply to other ZI types. The exception is Type D, which does lead to $\hat \mu^{(k)}=\bar y^{(k)}$, as we discuss below. 

The interpretation of Figure~\ref{fig:trajan_pit0_pi0} is assisted by the following observation. Optimisation overall at $(\hat {\lambda}_i,\hat\gamma )$ can be seen as optimisation for $\gamma$ (at $\hat\gamma$) given values of $\hat {\lambda}_i$, and for the $\lambda_i$ (at $\hat {\lambda}_i$) for given $\hat\gamma$. (We make no statement, here, as to the efficiency of an algorithm based on this; and any such algorithm should be distinguished from the EM algorithm widely used in ZI Type C, which uses binary regression of \emph{latent} binary variables.) In particular, for given $\hat {\lambda}_i$, the maximum likelihood estimation of $\gamma$ focuses exclusively on $\ell_y^0(\pit_0(\hat{\pi}_0,\gamma))$. That is to say, at the MLE, $\hat \gamma$ is that parameter which best fits the function $\pit_{0i}(\hat{\pi}_{0i}, \gamma)$ to observations $\Iyi$ modelled as $E[\IYi]$, which we denote as $\pit_{0i}$. Thus, $\hat \gamma$ may be considered (at the MLE) as the fitted parameter in the binary regression of the observables $\Iyi$ on $g(\hat \pi_{0i})$ using a very particular link function for $E[\IYi]$ with one unknown coefficient, that is $\gamma + g(\pit_{0i})$. A trivial example of this is that the MLE $\hat \gamma = p_0$ is, for the Trajan analysis, a weighted average of the eight separate values $p_0^{(k)}$. There being but one parameter in this function, the ultimate best fit is likely to be dominated by its large scale properties, being here the behaviour of the function at both large and small values of $\pi_0$; for Type A the functions in Figure~\ref{fig:trajan_pit0_pi0} are thus relatively inflexible.

For these data, Figure~\ref{fig:trajan_pit0_pi0} make it clear that ZI types A and C are evidently similar, and similarly inappropriate.  In contrast, the ZI types B and D are clearly more appropriate. More subtly, the data design -- resulting in large means and low probabilities of zero -- are challenging for the functions offered by types A and C, which offer little flexibility in this part of the unit square, unlike types B and D. Other data sets may of course be challenging for types B and D.

The second form for $\pit_0$ (Equation \eqref{inf2}) is of particular importance for Type D when the base distribution $\pi_y(\lambda)$ is, like the Poisson, within the Exponential Family. For the Trajan analysis, it highlights an important distinction between the models corresponding to (here similar) types B and D. In particular, it explains why, in Figure~\ref{fig:trajan_14panels}, pmfs for the Poisson, NB-quad and ZI Type D with Poisson base both lead to the same estimators of the cell expected values. ZI Type D may thus be a useful addition to the ZI types.

The key point in Equation \eqref{inf2} is that, for Type D, the odds ratio for $\pit_0$ and $\pi_0$ is taken to be constant, denoted $e^{\gamma}$. As a consequence, the parameter $\gamma$ can have a particular significance, when $\pi_y(\lambda)$ is in the (one parameter) Exponential Family and can thus be written as $\log(\pi_y) = y \eta(\lambda) - A(\lambda) + c(y)$; for the Poisson the natural parameter is $\eta = \log(\lambda)$. Firstly, defining $c(y)$ such that $c(0) = 0$, it is clear that the cumulant function $A(\lambda)=-\log(\pi_0)$, being $\lambda$ for the Poisson; and further, from the properties of the Exponential Family, we have $E[Y] = A^{\prime}_{\eta}(\lambda, \gamma)$, being $\lambda$ in the Poisson case. From \eqref{inf2} we thus have 
\begin{equation}
\log(\pit_y)  =  y \eta(\lambda) +\Iy\gamma  - \tilde A(\lambda, \gamma) + c(y)
\end{equation}
where $\tilde A(\lambda, \gamma) = \gamma - \log(\pit_0)$, which plays the same role as $A(\lambda)$ for $\pi_0$. The implication is that $\pit_y$ is in the two-parameter Exponential Family, inheriting this property from the $\pi_y$. The natural parameters are $(\gamma, \eta)$.

Further, the log-likelihood in Equation \eqref{logL} includes a decomposition into the sum of two terms, being $\ell^{0}_{y}(\gamma)$ and $\ell^{+}_{y}(\lambda)$. And indeed, if $\pi(y)$ is in the $m$-parameter family, then $\pit_y$ is in the $(m+1)$-parameter family. It follows immediately that sufficient statistics for an iid sample $y_i; i=1, \ldots n$ from $\pit_y$ are $\left(\sum y_i, \sum \Iyi\right)$, being equivalent to $(\bar y, p_0)$ and having expected values already known here to be $(\mu, \pit_0)$, recalling that $\mu = \rho \lambda$. This result is also available in general from differentiating $\tilde A(\lambda, \gamma)$. And, following the observation above on the equivalence, in the iid case, of all two parameter ZI types, $\bar y$ and $p_0$ are unbiased MLE estimators of $\mu$ and $\pit_0$, as already shown above, via the more cumbersome Type A arguments. This is the reason why ZI Type D (with Poisson base) shares this property with the usual Poisson and NB-quad models. And further, we can immediately assert that ZI Type D with NB-quad as base will also share this property.

But for the Trajan model, the decomposition in Equation \eqref{logL} extends to nine terms. Formally, with independent Poisson pmfs modelling each cell in the Trajan data, we can write this as an eight-parameter Exponential Family distribution. The log-likelihood for ZI Type D thus decomposes into the sum of nine terms, $\sum \ell_{\Iyi}(\gamma)$ and eight terms $\sum \ell_{y_i}(\lambda^{(k)})$. And thus the MLEs are $\hat \gamma=p_0$ (as above) and $\hat \mu^{(k)} = \bar y^{(k)}$. This differs (in respect of $\hat \mu^{(k)}$) from Type A above, and in fact coincides with the neutral (no ZI) Poisson mean for which $E[Y] = \lambda = \mu$ for each cell, as shown in Figure~\ref{fig:trajan_14panels}. The same reasoning applies to (neutral) NB-quad estimates and indeed to ZI Type D with NB-quad base (not shown here for brevity).

In closing this section we remark that arguments can be made for an alternative parameterisation of $\pit_0$ via $\mu$ (and thus of $\pit_y$), writing it instead as $\pit_0(\mu, \gamma)$, a function of $\pi_0(\lambda(\mu, \gamma))$. This might facilitate comparison with OD generalisations of the Poisson as base, where the parameter $\mu$ refers to $E[Y]$ in both base and generalisation. It would also have impact on Figure~\ref{fig:trajan_pit0_pi0}. But firstly, $\lambda(\mu, \gamma)$ is not generally available in closed form; and secondly, the arguments above for Type D show that the natural parameters -- in at least one theoretical sense -- are indeed (functions of) $\lambda$ and $\gamma$. There is at least one exception to this, provided by Type C; for there $\mu = \rho \lambda$, where $\rho=e^{\gamma}$ is constant, and we can write $\pit_0 = (1-\rho) + \rho \pi_0(\rho^{-1}\mu) =  (1-\rho) + \rho (\pi_0)^{\rho^{-1}}$ in the Poisson case, being no longer linear in $\pi_0$. Further, the importance of closed-form expressions will not be a barrier in some circumstances, such as diagnostics -- which are for discussion elsewhere.

\section{Zero-Inflation via Over-Dispersion}
\label{sec:ZI_via_NB}

As we have already noted, over-dispersion models, such as the negative binomial, have often been used for the analysis of count data with excess zeros. Indeed, some authors have suggested that in practice such OD models may be sufficient with no need to explicitly model the excess zeros. In Section~\ref{Sec:motiv_ex} we considered the use of particular negative binomial models and here we provide a little more detail. Of course the negative binomial family is just one form of OD extension of the Poisson  distribution, albeit the most widely studied and used. General mixing of a Poisson with any distribution (also often referred to as compounding) leads to OD count distributions. Other generalisations include weighted Poisson models \citep[see][]{DelCastillo1998} extending the original idea of size-weighting of counts in \citet{Rao1965}. This general family of models includes the currently popular COM-Poisson distribution \citep{Shmueli2005,Sellers2010,RibeiroJr2020} and unlike the mixture-based over-dispersion models this (and other members of the family) can also accommodate underdispersion. Other classes of extended Poisson models that can handle both over and underdispersion, include the Generalised Poisson \citep{Consul1973}, the Gamma-count \citep{Zeviani2014} based on using gamma-distributed inter-arrival times rather than the exponential times as in a Poisson process, and the discrete Weibull distribution \citep{Luyts2019} as an example of the general approach of discretisation of a continuous random variable. While OD versions of these can be used to model excess zeros, the combination of these models with explicit ZI also provides the possibility of modelling excess zeros where the non-zero counts are under-dispersed.

In many OD models not only do we have  $\mbox{Var}(Y) > \mbox{E}[Y]$, but also, often as a consequence,  $\pi_0(\mu, \phi) > \pi^P_0$. Simply stated, in general, over-dispersion puts more weight in the tails of a distribution, but as a count random variable $Y$ is constrained below by zero, this  extra weight in the lower tail accumulates on low values of $Y$ and, in particular, on $Y=0$. As such these models can provide an alternative approach to the apparent problem of excess zeros in data. As this is an indirect consequence of the OD model we call this \emph{implicit} ZI. \citet{Puig2006} provide a theoretical treatment of the circumstances under which OD induces ZI, characterising the conditions under which this is necessary. They use it to also contrast several count distributions; one of these is the Negative Binomial (NB), which here we use as an exemplar of OD models. Interestingly, as we develop below, the details depend not only on the OD distribution, but also on its parameterisation. In Figure~\ref{fig:trajan_pit0_pi0}, we show the implied ZI behaviour of these models for comparison with that of the explicit ZI models.

\subsection*{Negative Binomial}

We focus on two re-parameterisations of the Negative Binomial with pmf $$\pi_y(\mu,k) = \frac{\Gamma(k+y)}{y! \Gamma(k)}\left(\frac{k}{k+\mu} \right)^k \left(\frac{\mu}{k+\mu} \right)^y$$ which has $\mbox{E}[Y]=\mu$ and $\mbox{Var}(Y)= \mu +\mu^2/k$ and $\pi_0= \left(\frac{k}{k+\mu} \right)^k$. The most frequently used, to which we refer as NB-quad, has $k=\phi^{-1}$,  $\mbox{Var}(Y)= \mu +\phi \mu^2$ and $\pi_0= \left(1+\phi \mu \right)^{-\phi^{-1}}$. An alternative, NB-lin, has $k = \mu \phi^{-1}$, with $\mbox{Var}(Y)= \mu(1+ \phi)$ and $\pi_0=\left((1+\phi)^{\phi^{-1}}\right)^{-\mu}$. The Poisson is a special case of both, corresponding to $\phi \to 0$. There are in fact other versions of the Negative Binomial, generically NB-P \citep{Gurmu1996}, having $\mbox{Var}(Y)=\mu(1+\phi \mu^{p-1})$ with a power relationship for the over-dispersion index. Effectively this introduces a third parameter; but we do not pursue this option.

These two versions of the NB are included in Figure~\ref{fig:trajan_pit0_pi0}, where, as discussed in Section~3, the extended model zero probabilities $\pit_0$ are plotted against the Poisson base $\pi_0$, augmented by the observed and model zero probabilities for the Trajan data. Here we have expressed  $\pit_0(\mu, \phi)$ as a function of $\pi_0^P$ via $\mu = -\log \left(\pi_0^P\right)$. Figure~\ref{fig1_theoretical} also illustrates this for the two different parameterisations of the Negative Binomial distribution. In constructing this theoretical plot we have fixed the dispersion parameter $\phi$ so that the curves pass through the same common $(\pi_0,\pit_0)$ point $(0.2,0.4)$. For both, the induced ZI, characterised by $\pit_0(\pi^P_0, \phi)$ is such that $\pit_0 \to 0$ as $\pi_0^P \to 0$ or equivalently as $\mu \to \infty$, or equivalently as $\pi_0^P \to 0$, as with the ZI of types B and D.  More specifically, for NB-lin, from  $\log(\pit_0(\pi_0^P, \phi)) = \phi^{-1}\log(1+\phi) \log(\pi_0^P)$, if we set $e^{\gamma} = \phi^{-1}\log(1+\phi)$, the ZI induced by NB-lin is \emph{exactly} as Type B for all $\pi_0^P$ and thus for all $\mu$. Also, except for large $\mu$ (small $\pi_0^P$), this induced ZI  bears some resemblance to Type C, the classic mixture model of ZI. Further, we see that, apart from the behaviour for very small $\pi_0^P$ (ie for $\pi_0^P$ above the `elbow') the ZI function $\pi_0(\pi^P_0, \phi)$ for NB-quad is even more similar to that for Type C. This is essentially because here this function has unit slope both at the elbow and, unlike NB-lin, as $\pi_0^P \to 1$ (equivalently $\mu \to 0$). 

But recall that, despite ZI Type B and NB-lin having parameters $\gamma$ and $\phi$  such that they have the same probabilities of zero for all $\mu$, they are quite different distributions and have different variance functions. Although there are similarities in behaviour between the NB-quad and Type C models their probability structures are different and, for example, $\pit^C_y$ can be bi-modal. The point is that the over-dispersion in the Type C model is centred solely on the zero probability, whereas in the NB-quad the extra dispersion is smoothly spread across the whole range of $y$-values. The central challenge is that with only two parameters available, correspondence cannot be obtained on three important aspects: mean, variance and the probability of zero.

\subsection*{Simple Inference for the Negative Binomial}

Inference is well established for the NB model. But there is a subtlety, even in the iid case. Moments based estimators of $(\mu, \phi)$ will necessarily match the sample mean and variance; and thus not match $p_0$, the observed frequency of zeros, especially if this is large. But other, more widely used, estimators will, especially for large $p_0$, result in a $\hat \phi$ that (approximately) matches the fitted $\pi_0(\hat \mu, \hat \phi)$ by overestimating the variance. The theory is more straightforward for NB-quad.

NB-quad is a member of the Exponential Dispersion family. Here, in the iid case, the sample mean is the MLE for $\mu$; while the MLE for the dispersion parameter $\phi$ requires an iterative solution of the score equation. However, the joint ML estimators do have the nice property of being asymptotically independent \citep[see][]{Lawless1987}. Other estimators used for $\phi$ include a moment estimator (this can be obtained for more general models by equating the generalized Pearson chi-square statistic to the degrees of freedom). Historically, a  zero-frequency based estimator has also been used and, as for the explicit ZI models, it  results in the  zero-frequency being fitted perfectly; however, for data generated by an explicit ZI Poisson process, it does this by overestimating the variance with a larger fitted $\phi$ value. Interestingly in this ZI setting the MLE lies between the moment and the zero-frequency estimators, giving both a fitted variance $\mbox{Var}_{\pi_0(\hat \mu, \hat \phi)}$ that is larger than the observed sample variance, and a fitted $\pi_0(\hat \mu, \hat \phi)$ that is closer to the observed $p_0$ than might be expected, at the price of an extended fitted upper tail. However, in applications to data with excess zeros, there is typically also some OD in the non-zero counts and so this inflated estimated value of $\phi$ may simultaneously capture both excess zeros and additional dispersion and lead to an adequate and parsimonious fit. 

This behaviour perhaps explains the findings in \citet{Warton2005}, who provides comparisons between different implicit and explicit ZI models fitted to multivariate abundance data from 20 datasets. His results show that the NB-quad model is superior to ZIP and ZINB in terms of AIC, when looking at an average over 1672 count variables. He comments that these abundances do not have extra zeros when compared to the NB-quad distribution, and are likely to have arisen from NB-quad distributions with small means. 

The NB-lin is not in the Exponential Family, even for a known value of $\phi$, and  is inferentially less convenient. Of course, in the iid case the NB-lin is simply a reparameterisation of the NB-quad, but for fixed values of the dispersion parameters they behave differently as $\mu$ varies, see Table~\ref{tab:functions} and Figure~\ref{fig1_theoretical}, and these differences potentially become apparent in the regression model setting.

\section{R Packages for Fitting ZI Models}
\label{Sec:packages}

In this section we describe the use and background of various commonly-used R packages for modelling zero inflation. There are many packages listed on the R package web page which purport to perform zero-inflated regression, but many of these apply to specific data types or constraints (e.g. hidden Markov models, monotonic zero inflation, compositional data), apply to a particular application area, or deal in continuous rather than count data (the focus of this paper). Instead we focus on 4 main packages: \texttt{zic}, \texttt{pscl}, \texttt{VGAM}, and \texttt{gamlss}. A further popular package is \texttt{COZIGAM} but this package has been archived from CRAN and has not been updated since 2012. 

The purpose of this section is to showcase the wide array of possibilities currently available in R for zero inflated/adjusted modelling and provide brief examples of their use on our example data to enable those new to the field to pick up on the salient features. Table~\ref{tab_zi_models} shows a summary of these packages and an overview of their features and differences.

All of the aforementioned packages are Frequentist in their inferential approach with the exception of \texttt{zic}. A number of packages exist which potentially allow for Bayesian zero-inflated and zero-adjusted models to be fitted, such as JAGS \citep{Plummer2003}, Stan \citep{Team2014}, INLA \citep{rue2009approximate}, and Nimble \citep{DeValpine2017}. However, these require considerable extra coding experience (and often many more lines of code) so we do not review them here. A particular key issue is that many of these packages support only a small set of probability distributions, so that extra distributions have to be added by hand (or hack) and often lack vectorisation speed-ups. For those wishing to persist in learning one or more of these packages (which we would recommend), we suggest first consulting the manual to determine whether the chosen custom probability distributions are included by default. 

The \texttt{zic} package \citep{Jochmann2013} uses Markov chain Monte Carlo (MCMC) to produce posterior distributions of zero-inflated (though not hurdle) models with covariates. The covariates are assumed to be common across both the Poisson likelihood and the binary zero-inflation component. Only a Poisson likelihood is supported, though the model includes latent effects to account for over-dispersion to create, marginally, a Poisson log-normal likelihood. In addition the package has methods to run a stochastic search variable selection (SVSS), a method which removes the need to choose between model structures. 

R package \texttt{pscl} \citep{Zeileis2008} contains a very wide array of both hurdle and zero-inflated models. A key feature is the possibility of selecting different distribution types for $\pi_y$ and $\pi_0$. Three different distributions are permitted for the counts (Poisson, negative binomial, and geometric), and three types of zero-inflation (binomial, Poisson, negative binomial, and geometric). These latter distributions are censored at value 1 to produce the required hurdle effect. For standard zero inflation models only binomial (i.e. Bernoulli) inflation is permitted. Link functions can also be changed (for both hurdle and zero-inflated) though again some restrictions apply. 

The \texttt{VGAM} package interface \citep{Yee2015} is very similar to that of the standard R base \texttt{lm} and \texttt{glm} functions whilst adding in multivariate components (via constraint matrices) which do not apply to the zero inflated models implemented. Exactly as with \texttt{glm}, the \texttt{vglm} and \texttt{vgam} functions take a \texttt{formula} and \texttt{family} argument, the latter of which has a variety of zero-inflated options, including zero inflated Poisson, negative binomial, and geometric, and similarly zero-altered (two-stage/hurdle) versions. We found the output of these models to be somewhat confusing, since the tabular output displays multiple intercepts (corresponding seemingly to the different linear predictors). 

The \texttt{gamlss} package contains the richest set of zero-inflated and zero-altered (two-stage/hurdle) models, including a large number of zero-inflated continuous (and bounded) likelihood distributions. The count distributions supported include binomial, beta-binomial, negative binomial, negative beta-binomial, logarithmic, Poisson, and Zipf. The \texttt{gamlss} package in particular allows for the modelling of multiple different parameters in a distribution, each with their own parametric relationship, but none of the examples we found in the package used this feature.

Finally, we provide unified code to fit ZI types A, B, C and D (and reproduce all results obtained from analysing the Trajan data) at \url{www.github.com/andrewcparnell/ZI_review}.

\section{Conclusions}

In this paper, we have only examined the basic underpinnings of regression in the presence of excess zeros in univariate count data. Our first contribution to this is a novel perspective on the modelling of excess zeros in count data regression. This is primarily in the presentation of an approach which we have referred as explicit ZI.  The modelling issue, as presented here, is simply the choice of link function, and the consequent estimation of the sole parameter. Four such simple options are presented; others and extensions to two parameter variations are possible.  The two approaches discussed -- explicit ZI and implicit ZI via the use of distributions such as the NB -- may of course be combined, and all parameters may be modelled via covariates. 

It could be argued that the real difficulty is the embarrassment of choice. The estimation of the parameters is no longer a challenge (for univariate counts), given modern computing algorithms; nor is the identification of the \emph{best}, given a metric such as the AIC, a metric which may not be natural for some users. The real challenge is in fact the identification of the most \emph{useful} model given the vagaries of that term, and the ubiquity of potential outliers within all data sets. 

Our approach to the current paradigm of explicit ZI models  contrasts with that of data generating mechanisms that involve latent variables. Of course, as pointed out by \cite{Colin2013}, p.147: ``The latent class interpretation is not essential ... As such the approach is an alternative to non-parametric estimation." However, from the almost invariable discussion in  papers in the applied literature, it does seem that the user community feels obliged to explain the mixture interpretation. This can be useful, of course; but sometimes it can involve shoehorning. Nevertheless mixtures defined by latent variables can be a fruitful theoretical avenue for defining models for use with multivariate counts. Indeed this was the purpose of \cite{Salter-Townshend2012}. 

The practical benefit of the new perspective in univariate regression may not in fact be new models.  The non-parametric perspective may yield new diagnostics. It may even help to establish a-priori evidence of excess zeros. Current practice usually involves marginalisation over all covariates. An empirical approach would involve, as a first step, a comparison between observed $\Iyi$ and the fitted probabilities of zero under a default `base' model, generically $\hat \pi_{0i} = \pi_0\left(\hat \mu_i, \hat \phi_i\right)$. This comparison could simply regress the binary $\Iyi$ -- non-parametrically, using splines, for example -- on functions including $\log\left( \hat \pi_{0i}\right)$ and $\log\left( 1- \hat \pi_{0i}\right)$, yielding values $\hat \pit_{0i}$.  Any departure from the unit line $\pit_{0i} = \pi_{0i}$ would constitute a priori evidence of excess zeros with respect to the fitted base model. 

Furthermore such a plot might serve, post-hoc, as a diagnostic for a fitted ZI model, by regressing $\Iyi$ on functions of fitted $\pit_{0i}$. We remark that the current literature on count data regression in general -- and on excess zeros in particular -- seems sadly lacking in criticism that most users would deem to be constructive. Portmanteau statistics based on Information Criteria do not always have such a constructive interpretation. Model details can be dominated by extreme values; and this is particularly so for models that admit very large values for the variance. But IC statistics do not readily identify such data points.

In circumstances where it is deemed to be important to choose amongst alternative models for the excess zeros, such regressions highlight the importance of the design. For without data corresponding to very large $\mu_i$ (and thus small $\pi_{0i}$) it will be impossible to distinguish ZI models of Type B, C or D; and difficult to distinguish from the OD approach. 
In practice, of course, this will be further complicated if - as is common - the parameters $\gamma$ and/or $\phi$ are themselves modelled via covariates. But the fact that this is common may reflect poor diagnostics (or indeed over-enthusiasm).

Our second contribution is to make more explicit the parallels between the OD and ZI approaches to modelling excess zeros. But we have not resolved the choice between the explicit (ZI) and implicit (OD) approaches, when the base is the Poisson. On the contrary, we see that, from the point of view of the excess zeros, one ZI model -- Type B -- behaves exactly like the NB-lin; and NB-quad is not unlike the (dominant) ZI Type C. The difference between the latter two is only apparent for very large $\mu$, and only carefully designed data will differentiate them. They \emph{do} differ as regard the mean-variance relationship, but only in detail for both exhibit a quadratic relationship; and this too will only be apparent at very large $\mu$. In addition, for iid data, technical details of the usual inference procedures for NB-quad do little to help differentiate the very different models. Further, for the more typical case of regression, these details can only become more challenging. What is clear is that any attempt to discriminate between explicit and implicit ZI models will require a rich and varied dataset and diagnostics focussing jointly on both fitted zero-probability and upper tail behaviour. This discussion suggests that  the burgeoning literature on other over-dispersed generalisations of the Poisson -- and on zero-inflating them -- may itself be premature. As a final contribution, Table 1 offers an alternative to the confusing nomenclature that has developed. 

\section*{Acknowledgements}
Andrew Parnell’s work was supported by: a Science Foundation Ireland Career Development Award (17/CDA/4695); an investigator award (16/IA/4520); a Marine Research Programme funded by the Irish Government, co-financed by the European Regional Development Fund (Grant-Aid Agreement No. PBA/CC/18/01); European Union’s Horizon 2020 research and innovation programme under grant agreement No 818144; and SFI Research Centre awards 16/RC/3872 and 12/RC/2289\_P2. For the purpose of Open Access, the authors have applied a CC BY public copyright licence to any Author Accepted Manuscript version arising from this submission.

\bibliography{references.bib}

\newpage

\newpage

\newpage

\newpage

\renewcommand{\arraystretch}{1.5}
\begin{table}
	\centering
	\caption{A typology of zero inflation and over-dispersion. Our four types and their common names are shown alongside their ZI behaviour with respect to the Poisson base.}
	\vspace{.5cm}
	\begin{tabular}{ll}
		\hline \hline
		Type: Common names & Explicit ZI function \\[12pt]
		\hline
		\hline
		A: Basic Hurdle, Zero-Altered, Two-stage & $\logit(\pit_0) = \gamma$ \\[12pt]
		\hline
		B: Poisson Hurdle, Zero-Altered, Two-stage & $\log(-\log(\pit_0))= \gamma +\log(-\log(\pi_0))$ \\[12pt]
		\hline
		C: ZIP, Mixture, Lambert's mixture & $\log(1-\pit_0)= \gamma + \log(1-\pi_0), \gamma \le 0 $ \\[12pt]
		\hline
		D: Logistic ZI (New -- this paper) & $\logit(\pit_0)=\gamma + \logit(\pi_0)$ \\[12pt]
		\hline
		\hline
		OD dist & Implicit ZI function \\[12pt]
		\hline
		\hline
		NB-lin &  $\log(\pit_0)= \phi^{-1}\log(1+\phi)\, \log(\pi_0^P)$\\
		$\mbox{Var}(Y)=\mu+\phi\mu$& \\[12pt]
		\hline
		NB-quad & $\log(\pit_0)=-\phi^{-1} \log\left[1-\phi\log(\pi_0^P)\right]$ \\
		$\mbox{Var}(Y)=\mu+\phi\mu^2$&\\[12pt]
		\hline 
		\hline
	\end{tabular}
	\label{tab:functions}
\end{table}

\newpage
\renewcommand{\arraystretch}{1}

\begin{table}[htb]
	\caption{A list of R packages we use to demonstrate zero inflated regression modelling}
	\begin{tabular}{llp{5cm}p{6cm}}
		\hline
		Name & Approach & Likelihoods supported & Covariate types supported \\
		\hline
		\texttt{zic} & Bayesian & Poisson (and Poisson log-normal) & Restricted to be the same in both regular and zero-inflated component. Also includes variable selection \\
		\hline
		\texttt{pscl} & Frequentist & Poisson, negative binomial, and geometric (count distribution, and binomial, Poisson, negative binomial, and geometric (zero-inflated distributions; hurdle only) & Restricted to be the same in both regular and zero-inflated component \\ 
		\hline
		\texttt{VGAM} & Frequentist & Poisson, negative binomial, and geometric (Zero-inflated and hurdle) & Zero inflation components do not vary by covariate, but formulae allow for spline and other GAM-type relationships \\
		\hline
		\texttt{gamlss} & Frequentist & At least 12 different types (zero inflated and adjusted) & Has the ability to model each parameter in a distribution separately using, e.g. the GAM framework \\
		\hline
	\end{tabular}
	\label{tab_zi_models}
\end{table}

\end{document}